\def\rfr#1{eq. (\ref{#1})}

\def\bar{\begin{eqnarray}}
\def\ear{\end{eqnarray}}
\def\bb{\bibitem}
\def\eqi{\begin{equation}}
\def\eqf{\end{equation}}
\def\eqia{\begin{eqnarray}}
\def\eqfa{\end{eqnarray}}
\def\rp#1#2{{#1\over#2}}
\def\ct#1{\cite{#1}}
\def\lb#1{\label{#1}}

\def\oc2{$\mathcal{O}(c^{-2})$}

\documentclass[11pt]{article}
\usepackage{amsmath,amsthm,amscd,amssymb}
\usepackage{latexsym}
\usepackage{graphicx,epsfig}

\begin{document}

\noindent{\bf \LARGE{Model-independent test of spatial variations
of the Newtonian gravitational constant in some extrasolar
planetary systems}}
\\
\\
\\
{Lorenzo Iorio}\\
{\it Viale Unit$\grave{a}$ di Italia 68, 70125\\Bari, Italy
\\tel./fax 0039 080 5443144
\\e-mail: lorenzo.iorio@libero.it}

\begin{abstract}
In this paper we directly constrain possible spatial variations of
the Newtonian gravitational constant $G$ over ranges $\approx
0.01-5$ AU in various extrasolar multi-planet systems. By means of
the third Kepler's law we determine the quantity $\Gamma_{\rm
XY}=G_{\rm X}/G_{\rm Y}$ for each couple of planets X and Y
located at different distances from their parent star: deviations
of the measured values of $\Gamma$ from unity would signal
variations of $G$. The obtained results for $\eta=1-\Gamma$ are
found to be well compatible with zero within the experimental
errors ($\eta/\delta\eta\approx 0.2-0.3$). A comparison with an
analogous test previously performed in our Solar System is made.
\end{abstract}

Key words: stars: planetary systems-stars:
exoplanets-celestial mechanics-ephemerides-gravitation\\

PACS: 04.80.-y, 95.10.Ce, 04.80.Cc, 97.82.-j\\

\section{Introduction}
More than 200 exoplanets have been so far discovered with
different techniques such as radial velocity\footnote{It allows
for the most accurate ephemerides.}, microlensing\footnote{Until
now there are only four planets discovered with such method: their
orbital parameters are rather poorly determined for our
purposes.}, direct imaging\footnote{Until now there are only four
planets discovered with such method: only their masses and
semimajor axes are available for them.}. For an update list see on
the WEB http://exoplanet.eu/. One of the most important orbital
parameters phenomenologically measured in such systems with
usually great accuracy is the sidereal orbital period: since it is
affected by many Newtonian and non-Newtonian effects, it may be
used, in principle, as an important probe to measure or constrain
such  features within the experimental errors. It is our intention
to use some of the best determined multiple planetary systems in
order to put on the test a key feature of the gravitational
interaction, i.e. the possibility that the Newtonian gravitational
constant $G$ may vary with distance.
\subsection{Overview of the considered multi-planet systems}
$\upsilon$ And (HD 9826)  is a 2.41 Gyr old  main sequence star of
spectral type F8V located at a distance of 13.47 pc from us with
RA (J2000) 01 36 47.843 and DEC (J2000) +41 24 19.65  \ct{Val05}.
Its mass and radius are $M=1.32$M$_{\odot}$ \ct{Val05} and
$R=(1.15\pm 0.15)$R$_{\odot}$ \ct{Pas01}, respectively. It harbors
a planetary system \ct{Nae04, But06} composed by three
Jupiter-type planets: the closest one orbits $\upsilon$ And at a
0.0595 AU distance, while the farthest one is at 2.54 AU from its
parent star. The relevant orbital parameters are in Table
\ref{UpsAnd}.
\begin{table}\caption{ Relevant parameters \ct{But06} of the three planets
$\upsilon$ And b, $\upsilon$  And c and $\upsilon$ And d.  $a$ is
the semimajor axis, in AU, $e$ is the eccentricity, $P$ is the
sidereal orbital period, in days, and $m\sin i$ is the minimum
planet's mass, in Jovian masses. The stellar mass and radius are
$M=1.32$M$_{\odot}$ \ct{Val05} and $R=(1.15\pm 0.15)$R$_{\odot}$
\ct{Pas01}, respectively. } \label{UpsAnd}

\begin{tabular}{lllll} \noalign{\hrule height 1.5pt}

Planet & $a$ (AU) & $e$ & $P$ (d) & $m\sin i$ (m$_{\rm Jup}$)\\
b & $0.0595\pm 0.0034$ & $0.023\pm 0.018$ & $4.617113\pm 0.000082$  & $0.687\pm 0.058$\\
c & $0.832\pm 0.048$ & $0.262\pm 0.021$ & $241.23\pm 0.30$ & $1.98\pm 0.17$\\
d & $2.54\pm 0.15$ & $0.258\pm 0.032$ & $1290.1\pm 8.4$ & $3.95\pm 0.33$\\
\hline

\noalign{\hrule height 1.5pt}
\end{tabular}

\end{table}

Another triple system \ct{San04, But06} is hosted by the 6.41 Gyr
old main sequence star $\mu$ Ara (HD 160691) at 15.28 pc from us
with RA (J2000) 17 44 08.703 and DEC (J2000) -51 50 02.59
\ct{Val05}. Its spectral type is G3 IV-V, its mass is
$M=1.15$M$_{\odot}$ \ct{Val05} and its radius is $R=(1.245\pm
0.255)$R$_{\odot}$ \ct{Pas01}. The relevant orbital parameters of
such a system, consisting of two Jupiter-type bodies and one
Neptune-like planet, are in Table \ref{MuAra}. The closest planet
orbits at 0.0924 AU from its parent star, while the farthest one
is at 3.78 AU from $\mu$ Ara.
\begin{table}\caption{ Relevant parameters of the three planets
$\mu$ Ara b \ct{But06}, $\mu$  Ara c \ct{But06} and $\mu$ Ara d
\ct{San04}. $a$ is the semimajor axis, in AU, $e$ is the
eccentricity, $P$ is the sidereal orbital period, in days, and
$m\sin i$ is the minimum planet's mass, in Jovian masses. The
stellar mass and radius are $M=1.15$M$_{\odot}$ \ct{Val05} and
$R=(1.245\pm 0.255)$R$_{\odot}$ \ct{Pas01}, respectively. }
\label{MuAra}

\begin{tabular}{lllll} \noalign{\hrule height 1.5pt}

Planet & $a$ (AU) & $e$ & $P$ (d) & $m\sin i$ (m$_{\rm Jup}$)\\
b & $1.510\pm 0.088$ & $0.271\pm 0.040$ & $630.0\pm 6.2$  & $1.67\pm 0.17$\\
c & $3.78\pm 0.25$ & $0.463\pm 0.053$ & $2490\pm 100$ & $1.18\pm 0.12$\\
d & $0.0924\pm 0.0053$ & $0.000\pm 0.020$ & $9.550\pm 0.030$ & $0.0471$\\
\hline

\noalign{\hrule height 1.5pt}
\end{tabular}

\end{table}

Three Neptune-type planets \ct{Lov06} have recently been
discovered around the 7 Gyr old main sequence star HD 69830
\ct{Lov06} at 12.6 pc from us with RA (J2000) 08 18 23 and DEC
(J2000) -12 37 55. Its spectral class is K0V, its mass is
$M=(0.86\pm 0.03)$M$_{\odot}$ and its radius is $R=(0895.\pm
0.005)$R$_{\odot}$. The orbital parameters of such system are in
Table \ref{3nep}.
\begin{table}\caption{ Relevant parameters \ct{Lov06} of the three planets
HD 69830 b, HD 69830 c and HD 69830 d. $a$ is the semimajor axis,
in AU, $e$ is the eccentricity, $P$ is the sidereal orbital
period, in days, and $m\sin i$ is the minimum planet's mass, in
Jovian masses. The stellar mass and radius are $M=(0.86\pm
0.03)$M$_{\odot}$ and $R=(0.895\pm 0.005)$R$_{\odot}$,
respectively. } \label{3nep}

\begin{tabular}{lllll} \noalign{\hrule height 1.5pt}

Planet & $a$ (AU) & $e$ & $P$ (d) & $m\sin i$ (m$_{\rm Jup}$)\\
b & $0.0785$ & $0.1\pm 0.04$ & $8.667\pm 0.003$  & $0.033$\\
c & $0.186$ & $0.13\pm 0.06$ & $31.56\pm 0.04$ & $0.038$\\
d & $0.63$ & $0.07\pm 0.07$ & $197\pm 3$ & $0.058$\\
\hline

\noalign{\hrule height 1.5pt}
\end{tabular}

\end{table}

A tetra-planetary system  \ct{McA04, Nae04} is hosted by the 5.5
Gyr old main sequence star 55 Cnc ($\rho^1$ Cnc, HD 75732) at 13.4
pc from us with RA (J2000) 08 52 35.811 and DEC (J2000) +28 19
50.95. Its spectral type is G8 V, its mass is $M=0.91$M$_{\odot}$
\ct{Val05} and its radius is $R=(1.245\pm 0.255)$R$_{\odot}$
\ct{Pas01}. Such a system is formed by three Jupiter-like planets
and one Neptune-type planet spanning the range $0.0377-5.97$ AU
from their parent star. See Table \ref{55Cnc}.
\begin{table}\caption{ Relevant parameters \ct{McA04} of the four planets
55 Cnc b, 55 Cnc c,  55 Cnc d and 55 Cnc e. $a$ is the semimajor
axis, in AU, $e$ is the eccentricity, $P$ is the sidereal orbital
period, in days, and $m\sin i$ is the minimum planet's mass, in
Jovian masses. The stellar mass and radius are $M=0.91$M$_{\odot}$
\ct{Val05} and $R=(0.6\pm 0.3)$R$_{\odot}$ \ct{Pas01},
respectively. } \label{55Cnc}

\begin{tabular}{lllll} \noalign{\hrule height 1.5pt}

Planet & $a$ (AU) & $e$  & $P$ (d) & $m\sin i$ (m$_{\rm Jup}$)\\
b & $0.1138\pm 0.0066$ & $0.01\pm 0.13$ & $14.652\pm 0.010$  & $0.833\pm 0.069$\\
c & $0.238\pm 0.014$ & $0.071\pm 0.012$ & $44.36\pm 0.25$ & $0.157\pm 0.020$\\
d & $5.97\pm 0.35$ & $0.091\pm 0.080$ & $5552\pm 78$ & $3.90\pm 0.33$\\
e & $0.0377\pm 0.0022$ & $0.09\pm 0.28$ & $2.7955\pm 0.0020$ & $0.0377\pm 0.0059$\\
\hline

\noalign{\hrule height 1.5pt}
\end{tabular}

\end{table}

As can be noted, the considered planetary systems, all discovered
with the radial velocity method, are hosted by main sequence,
Sun-like stars located at about $12-15$ pc from us.
\subsection{Aim of the paper}
Such extra-solar systems offer us different laboratories outside
our Solar System to perform direct and model-independent tests of
possible variations with distance of the Newtonian gravitational
constant $G$ over scales ranging from $\approx 0.01$ AU to
$\approx 5$ AU.
%
\section{Spatial variations of the Newtonian gravitational constant in the considered planetary systems}
The possibility that the Newtonian gravitational constant $G$ may
experience spatial variations is envisaged by many theoretical
frameworks dealing with generalized theories of gravity and
unified theories of basic physical interactions \ct{Des92, Mel94,
Mel96, Bar01}. In, e.g., the extended chaotic inflation scenario
proposed by Linde in \ct{Lin90} the values of the effective
gravitational constant in different parts of the universe may
differ from each other. Also scalar-tensor theories predict that
$G$ may spatially vary \ct{Cli05}. A scalar-tensor-vector-theory
which, among other things, predicts spatial variations of $G$
\ct{Bro06} is the one by Moffat \ct{Mof06}. Some non-perturbative
studies of quantum gravity suggest that the effective $G$ might
slowly increase with distance \ct{Ham05}; in cosmology, this may
work as an alternative to dark matter and be related to the
expansion acceleration. Spatial variations of $G$ are also
predicted in the frameworks of the Yukawa-like modifications of
Newtonian gravity \ct{Kra01, Fis98}, and MOND (MOdified Newtonian
Dynamics) \ct{Mil83, McG02, Fam05, Snd06, Zha06}.

Earth-based, laboratory-scale investigations of spatial variations
of $G$ can be found in \ct{Izm93, Ger02, Lon03}. Both laboratory
and astronomical tests can be found in \ct{Mik77, Hut81, Bur88}.
Constraints on variations of $G$ with scale from gravitational
lensing and the cosmic virial theorem are reported in \ct{Ber96}.
Large (cosmological)-scale bounds on spatial variations of $G$
have recently been placed by Barrow in \ct{Bar05}. Effects of
possible spatial variations of $G$ on the cosmic microwave
background are reported in \ct{Bou05}.

A useful approach to test spatial variations of $G$ in typical
astronomical scenarios as the planetary systems considered here is
the following one. According to the third Kepler Law, \eqi
G(a)=\left(\rp{2\pi}{P}\right)^2\rp{a^3}{M},\lb{kep}\eqf so that
for a generic pair of planets X and Y we can construct the ratio
 $\Gamma_{\rm XY}=G_{\rm X}/G_{\rm Y}$ as \eqi
\Gamma_{\rm XY}=\left(\rp{P_{\rm Y}}{P_{\rm
X}}\right)^2\left(\rp{a_{\rm X }}{a_{\rm
Y}}\right)^3\lb{gamma};\eqf deviations of such a quantity from
unity would reveal scale variations of $G$. Such an approach was
used  in our Solar System \ct{Tal88} by comparing $\Gamma$ to
unity as $\eta=1-\Gamma$ for those planets for which radar-ranging
measurements of their orbital radiuses existed. No evidence for
any anomalous behavior of the Newtonian gravitational constant
from the orbits of Mercury to that of Jupiter, i.e. in the range
$0.38-5$ AU, was found; all the determined values for $\eta$ were
less than $1$ sigma from zero, except for Venus, which was a $1.6$
sigma result.

Having at our disposal the phenomenologically measured  semimajor
axes and orbital periods of the $\upsilon$ And, $\mu$ Ara, HD
69830 and 55 Cnc planets, we can accurately map possible
deviations of $\Gamma$ from unity in each of such systems.  It is
important to note that the precision with which we presently know
the planets' periods allows us to neglect the corrections to the
simple Keplerian model of \rfr{kep} due to both the quadrupole
mass moments $J_2$ of the parent stars \ct{Ior06b} and the
post-Newtonian component of their gravity field of order
$\mathcal{O}(c^{-2})$ \ct{Sof89, Mash01}.
Table \ref{Gamma} shows our results for the variations of $G$ in
the $\upsilon$ And, $\mu$ Ara, HD 69830 and 55 Cnc systems.
\begin{table}\caption{$\upsilon$ And, $\mu$  Ara, HD 69830 and $55$ Cnc systems:
variation of the gravitational constant $G$ in the spatial regions
crossed by the planets. } \label{Gamma}

\begin{tabular}{lllllll} \noalign{\hrule height 1.5pt}

System & $\Gamma_{\rm bc}$ & $\Gamma_{\rm bd}$ & $\Gamma_{\rm cd}$ & $\Gamma_{\rm be}$ & $\Gamma_{\rm ce}$ & $\Gamma_{\rm de}$\\
$\upsilon$ And & $0.9\pm 0.3$ & $1.0\pm 0.4$ & $1.0\pm 0.4$&-&-&-\\
$\mu$ Ara & $0.9\pm 0.4$ & $1.0\pm 0.4$ & $1.0\pm 0.5$&-&-&-\\
HD 69830 & $0.997\pm 0.003$ & $0.99\pm 0.02$ & $1.00\pm 0.02$ &-&-&-\\
$55$ Cnc & $1.0 \pm 0.4$ & $0.9\pm 0.4$ &   $0.9\pm 0.4$   &  $1.0\pm 0.3$  & $1.0\pm 0.4$ & $0.9\pm 0.4$\\
\hline

\noalign{\hrule height 1.5pt}
\end{tabular}

\end{table}
%
%
%
%
%
%
%
%
%
%
%
%
%
%
%
%
%
%
%
%
%
%
%
%
%
%
%
%
%
%
The uncertainty was calculated as \eqi \delta\Gamma_{\rm XY
}\leq\delta\Gamma|_{P_{\rm Y}} + \delta\Gamma|_{P_{\rm X}} +
\delta\Gamma|_{a_{\rm X}} + \delta\Gamma|_{a_{\rm Y}},\eqf with
\begin{equation}\left\{\begin{array}{lll}
\delta\Gamma|_{P_{\rm Y}} \leq 2\rp{P_{\rm Y}}{P^2_{\rm X}}\left(\rp{a_{\rm X}}{a_{\rm Y}}\right)^3\delta P_{\rm Y},\\\\
\delta\Gamma|_{P_{\rm X}} \leq 2 \rp{P^2_{\rm Y}}{P^3_{\rm
X}}\left(\rp{a_{\rm X}}{a_{\rm Y}}\right)^3\delta P_{\rm X},\\\\
\delta\Gamma|_{a_{\rm X}}\leq 3\left(\rp{P_{\rm Y}}{P_{\rm X}}\right)^2\rp{a^2_{\rm X}}{a^3_{\rm Y}}\ \delta a_{\rm X},\\\\
\delta\Gamma|_{a_{\rm X}}\leq 3\left(\rp{P_{\rm Y}}{P_{\rm
X}}\right)^2\rp{a^3_{\rm X}}{a^4_{\rm Y}}\ \delta a_{\rm
Y}.\lb{errori}
\end{array}\right.\end{equation}
In the case of  $\upsilon$ And we note that our result for
$\Gamma$ is accurate at $2.5-3$ sigma. For $\mu$ Ara the obtained
precision in determining $\Gamma$ is $2-2.5$ sigma. For HD 69830
the situation is much better
 because only the errors in the orbital periods were accounted
for, being those in the semimajor axes not released in \ct{Lov06}.
In the 55 Cnc system $\Gamma$ was measured at a $2.2-3.3$ sigma.
Reasoning in terms of deviations of the measured quantity $\Gamma$
from the expected value 1, i.e. by considering $\eta$, we see that
from Table \ref{gamma} it is possible to obtain figures for such a
quantity which are well compatible with zero differing from it by
only $0.2-0.3$ sigma or less. We note that in our test we used
four different planetary systems involving a total of twelve
planets; instead, in the Solar System test of \ct{Tal88} only four
planets were considered: for one of them, i.e. Venus, $\eta$ was
found to be compatible with 0 at 1.6 sigma level, while for
Mercury, Mars and Jupiter the agreement was at about 0.8, 0.3 and
0.5 sigma level, respectively. Thus, our test can, in general, be
considered more accurate than the one of \ct{Tal88}. Moreover,
when more data from the considered planetary systems will become
available and will be processed, it will be possible to further
improve the precision of such an extra-solar test.

The most general conclusion that can be drawn is that our results
for spatial variations of $G$ throughout the extension of the
considered planetary systems are compatible with zero within the
experimental errors. Our analysis relies only upon the third
Kepler law, being independent of any model of modified gravity
predicting spatial variations of $G$.

\section{Conclusions}
In this paper we constrained, in a model-independent way, spatial
variations of the Newtonian gravitational constant $G$ throughout
different extrasolar multi-planet systems extending from 0.01 AU
to $2-5$ AU and located at about $12-15$ pc from us. Our results
are compatible with zero within the experimental errors
($\eta/\delta\eta\approx 0.2-0.3$). With respect to what
previously done in our Solar System with Mercury, Venus, Mars and
Jupiter, here we used a total of 13 planets in four
independent planetary systems sampling a wider range of distances
from their central stars thanks to $\upsilon$ And b, $\mu$ Ara d,
HD 69830 b and 55 Cnc e at orbiting at about 0.01 AU from their
parent stars. The precision achieved is better than that of the
Solar System test and will be increased when more data from such
planetary systems will be collected and processed.


%
\section*{Acknowledgments}
I gratefully thank P. Salucci (SISSA) for important remarks on MOND

\end{document}